\documentclass[manuscript,screen,authorversion]{acmart}
\AtBeginDocument{%
  \providecommand\BibTeX{{%
    \normalfont B\kern-0.5em{\scshape i\kern-0.25em b}\kern-0.8em\TeX}}}

\setcopyright{rightsretained}
\copyrightyear{2023}
\acmYear{2023}
\acmDOI{XXXXXXX.XXXXXXX}

\usepackage{multirow}
\usepackage{booktabs}
\usepackage{caption}
\usepackage{subcaption}

\begin{document}

\title[Distributive Justice as the Foundational Premise of Fair ML]{Distributive Justice as the Foundational Premise of Fair ML: Unification, Extension, and Interpretation of Group Fairness Metrics}

\author{Joachim Baumann}
\authornote{Equal contribution.}
\email{baumann@ifi.uzh.ch}
\orcid{0000-0003-2019-4829}
\affiliation{%
  \institution{Zurich University of Applied Sciences, University of Zurich}
  \country{Switzerland}
}

\author{Corinna Hertweck}
\authornotemark[1]
\email{corinna.hertweck@zhaw.ch}
\orcid{0000-0002-7639-2771}
\affiliation{%
  \institution{Zurich University of Applied Sciences, University of Zurich}
  \country{Switzerland}
}

\author{Michele Loi}
\email{michele.loi@polimi.it}
\orcid{0000-0002-7053-4724}
\affiliation{%
  \institution{Polytechnic University of Milan}
  \country{Italy}
}

\author{Christoph Heitz}
\email{christoph.heitz@zhaw.ch}
\orcid{0000-0002-6683-4150}
\affiliation{%
  \institution{Zurich University of Applied Sciences}
  \country{Switzerland}
}

\begin{abstract}
Group fairness metrics are an established way of assessing the fairness of prediction-based decision-making systems. However, these metrics are still insufficiently linked to philosophical theories, and their moral meaning is often unclear. In this paper, we propose a comprehensive framework for group fairness metrics, which links them to more theories of distributive justice. The different group fairness metrics differ in their choices about how to measure the benefit or harm of a decision for the affected individuals, and what moral claims to benefits are assumed. Our unifying framework reveals the normative choices associated with standard group fairness metrics and allows an interpretation of their moral substance. In addition, this broader view provides a structure for the expansion of standard fairness metrics that we find in the literature. This expansion allows addressing several criticisms of standard group fairness metrics, specifically: (1) they are parity-based, i.e., they demand some form of equality between groups, which may sometimes be detrimental to marginalized groups; (2) they only compare decisions across groups but not the resulting consequences for these groups; and (3) the full breadth of the distributive justice literature is not sufficiently represented.
\end{abstract}


\maketitle

\section{Introduction}
\label{sec:introduction}

Supervised machine learning (ML) is increasingly used for prediction-based decision-making in various consequential applications, such as credit lending, school admission, and recruitment.
Research has shown that the use of algorithms for decision-making can reinforce existing biases or introduce new ones~\cite{Kleinberg2017HumanDecisionsMachinePredictions,10.2307/24758720}.
Consequently, fairness has emerged as an important desideratum for automated decision-making.
As many cases have shown, considering fairness explicitly is crucial in order to avoid disadvantages towards marginalized groups (see, e.g.,~\cite{machine-bias, markup2021, crawford2016artificial, buolamwini2018gender, br-retorio, obermeyer2019dissecting}).

Different measures have emerged in the algorithmic fairness literature for assessing unfairness in decision-making systems, many of which are in the category of so-called group fairness criteria.\footnote{Readers unfamiliar with group fairness may refer to~\cite{narayanan2018translation}, \cite{verma2018fairness}, and \cite[Chapter 3]{fairmlbook} for an overview of the topic, and to Appendix~\ref{sec:appendix-group-fairness} for a brief introduction of the most-discussed group fairness criteria.}
The concept of group fairness stands in contrast to approaches focusing on individuals, such as individual fairness~\cite{dwork2012fairness,Speicher2018individualandgroupfairness}, 
or counterfactual fairness~\cite{kusner2017counterfactual}.
This paper focuses on group fairness metrics. 

\citet{heidari2019moral} provide a unifying framework for these criteria. However, they only consider standard fairness criteria that demand equality between different socio-demographic groups, i.e., that are based on an egalitarian notion of distributive justice~\cite{binns2020apparent}.
They do not discuss non-egalitarian fairness criteria, which -- as we will see in Section \ref{sec:extension} -- can be relevant for the assessment of fairness.
\citet{kuppler2021distributive} find that 
``apparently, the fair machine learning literature has not taken full advantage of the rich and longstanding literature on distributive justice''~\cite[p.~17]{kuppler2021distributive}.
Our paper addresses this gap by building on extensions of standard group fairness criteria and linking them to the distributive justice literature, considering both egalitarian and non-egalitarian concepts.
We propose a generalized framework for assessing the fairness of decision systems, drawing on the concept of distributive justice. Based on this, we offer a generalized definition of group fairness, which includes the known group fairness criteria but significantly extends the space of group fairness. 

While decision systems are usually designed to optimize a certain goal for a decision maker, they also produce some benefit or harm for the affected individuals. 
On a societal scale, the repetitive application of the decision system leads to a \emph{distribution of benefit/harm} among different social groups. We study the question of group fairness by building on theories of distributive justice, which are concerned with the question of when such a distribution can be called just.
Our suggested framework consists of the following four components:
\begin{enumerate}
    \item \textbf{Utility of the decision subjects:} Defines how to measure the amount of benefit/harm for decision subjects.
    \item \textbf{Relevant groups:} Defines the social groups to be compared with respect to how much utility they receive.
    \item \textbf{Claim differentiator:} Defines the features which justify inequalities in the distribution of utility between individuals.
    \item \textbf{Pattern of justice:} Defines what constitutes a just distribution. 
\end{enumerate}
All four components represent normative choices about what constitutes justice or fairness and are built on existing work.
The four components as such are thus not novel but rather an established part of the literature on fairness metrics. The key novelty of our paper is the combination of these existing components into a comprehensive framework for group fairness metrics.
This is important for the following reasons:
\begin{itemize}
    \item \textbf{Unification:} We show that the most popular group fairness metrics can be interpreted as instantiations of our framework. Thus, the framework provides a unification of established group fairness metrics, interpreting them as different applications of a common general principle of distributive justice. 
    \item \textbf{Extension:} Our framework is built on a generalized definition of group fairness, which establishes the general structure of group fairness criteria and also suggests ways to diverge from established criteria. Therefore, the framework can be used to construct new criteria that are adapted to the context of the application.
    \item \textbf{Interpretation:} Each component in our framework is linked to particular aspects of the moral assessment of a decision-making system. When we interpret established group fairness criteria as special cases of our framework, we can thus explicate the assumptions that are implicitly embedded in these group fairness criteria. Thereby, we provide new insights into established group fairness criteria and make it easier to evaluate whether a fairness criterion is morally appropriate for a given context.
\end{itemize}

The paper is structured as follows: We first present existing literature on group fairness in Section~\ref{sec:related-work}. Specifically, we will discuss the limitations of standard group fairness metrics and how existing work has expanded on these standard metrics.
In Section~\ref{sec:framework}, we present our comprehensive framework for group fairness.
We focus on the mathematical formalization of different aspects of the distributive justice literature while keeping the review of the philosophical foundations short.
More details about the philosophical background can be found in the companion paper~\cite{Hertweck2023JusticeBasedFramework}.
Section~\ref{sec:relation-group-fairness} then demonstrates that standard group fairness metrics are special cases of our group fairness framework.
Next, in Section~\ref{sec:example}, we showcase the extensions of our framework compared to existing approaches using an example from the medical domain.
Finally, we discuss the implications of our framework and possible future work in Section~\ref{sec:discussion}.

\section{Related work}\label{sec:related-work}

Our work focuses on group fairness criteria.
The most popular group fairness criteria have been developed in the context of binary classification problems and are derived from the confusion matrix: Conditional probabilities such as the true positive rates or the positive predictive values are compared across groups. We refer to these as ``standard group fairness criteria'' (see Appendix~\ref{sec:appendix-group-fairness}). In this section, we first take a look at the limitations of the standard group fairness criteria and then discuss how they have been expanded in other works.

\subsection{Limitations of standard group fairness criteria}\label{sec:limitation-group-fairness}

There does not seem to be a clear consensus on what group fairness is and different terms have been used in the literature to describe the concept. To frame our understanding of the literature's current view on (standard) group fairness criteria, we refer to the following definitions:
\begin{enumerate}
\item ``Group fairness ensures some form of statistical parity (e.g. between positive outcomes, or errors) for members of different protected groups (e.g. gender or race)''~\cite[p.~514]{binns2020apparent} (based on \cite{dwork2012fairness}'s definition of statistical parity)
\item ``Different statistical fairness criteria all equalize some group-dependent statistical quantity across groups defined by the different settings of [the sensitive attribute] $A$''~\cite[Chapter 3]{fairmlbook}.
\end{enumerate}
These definitions show the following three common properties of standard group fairness metrics: (1) they consider multiple groups, (2) they compare averages over groups, and (3) they demand parity between these (what is referred to as \textit{egalitarianism}).%
\footnote{
Note that these definitions of group fairness also fall into the category of ``oblivious measures''~\cite[p.~3]{hardt2016equality} and ``fairness definitions from data alone''~\cite[p.~149]{mitchell2021algorithmic}, i.e., measures that only require access to the data of the decision-making system.
This stands in contrast to alternative concepts of fairness that ``incorporate additional context''~\cite[p.~154]{mitchell2021algorithmic} (such as individual fairness \cite{dwork2012fairness} or causal definitions of fairness \cite{kusner2017counterfactual}), which we do not consider here.
}
Standard group fairness metrics suffer from several limitations:

\paragraph{The ``leveling down objection''}

As shown by~\cite{hu2020fair}, enforcing equality can yield worse results for \emph{all} groups. This so-called ``leveling down objection'' is often brought forward to challenge egalitarianism in philosophical literature~\cite{parfit1995,crisp2003}: In a case in which equality requires us to worsen the outcomes for everyone, should we really demand equality or should we rather tolerate some inequalities?\footnote{One may argue that in a case where equality is not met, one should opt for, e.g., the collection of better data instead of worsening a group's utility. However, it cannot be ruled out that some form of de-biasing would still be necessary and could then worsen a group's utility. Societal inequalities, for example, persist even in ``better'' data, so there is no guarantee that equality can be achieved while keeping the same level of utility for all groups.} As criticized by \cite{feder2021emergent} and \cite{weerts2022does}, standard definitions of group fairness lack this differentiation as they always minimize inequality.

\paragraph{Focus on decisions instead of consequences}

As pointed out by \cite{hertweck2021moral} and \cite{weerts2022does}, standard fairness criteria like statistical parity or equality of opportunity focus on an equal distribution of favorable \emph{decisions} and not on the \emph{consequences} of these decisions. They assume ``that a `positive classification' output is an equally valuable outcome for everyone'' as pointed out in \cite[p.~491]{finocchiaro2021bridging}. Similarly, \cite{binns2018fairness} notes that these criteria ``[assume] a uniform valuation of decision outcomes across different populations'' \cite[p.~6]{binns2018fairness}, and highlights that this assumption does not always hold. 
This makes it difficult to use standard group fairness criteria for a moral assessment of unfairness. 
Moreover, parity-based criteria do not allow for unequal treatment, even if this may be desirable from a social justice perspective in certain cases~\cite{kasy2021fairness}.

\paragraph{Limited set of fairness definitions}

Standard group fairness metrics differ with respect to their underlying moral values \cite{selbst2019fairness,jacobs2021measurement}. As they are mathematically incompatible, one has to choose one over the others \cite{kleinberg2016inherent,chouldechova2017fair, fairmlbook,kleinberg2019discrimination,Wong2020}.
None of the standard group fairness criteria proposed in the algorithmic fairness literature might be morally appropriate in a given context (see the example in Section~\ref{sec:example}).

\subsection{Extension of standard group fairness criteria}\label{sec:extension}

New group fairness metrics have been suggested to overcome the limitations of standard group fairness metrics. Several works have taken a utility-based view of fairness to overcome the issue that standard fairness metrics do not consider the mapping of decisions to a benefit/harm for decision subjects.\footnote{Note that some of this literature uses the term ``welfare'' instead of ``utility'', which can be traced back to the different fields these works intersect with.}
\citet{finocchiaro2021bridging} point out that ``utilitarianism and normative economics have been extensively used in mechanism design to motivate using utility functions as a synonym for social welfare'' and suggest that machine learning could build on this through, for example, ``individual and group-level utilities'' \cite[p.~491]{finocchiaro2021bridging}.
Individual utilities have been used to define ``fairness behind a veil of ignorance'' \cite{heidari2018fairness}.
Several works conceive group-based notions of fairness that are centered on utility \cite{hu2020fair, ben2021protecting, hossain2020designing}. \cite{ben2021protecting} show that enforcing fairness criteria may harm marginalized groups if the wrong utility values are assumed. Similar to our findings, they also point out that standard group fairness metrics map to utility-based group fairness metrics.
However, they do not move beyond parity-based notions of fairness.

\cite{hossain2020designing} adapts the concept of envy-freeness to fair machine learning by ``requir[ing] that individuals in group $G$ [do] not prefer the classification given to individuals in group $\hat{G}$ (and not just the classification that would be given to them if the classifier for group $\hat{G}$ were used for them)'' \cite[p.~2]{hossain2020designing}. They show how standard group fairness criteria can be mapped to this interpretation of group-level envy-freeness. While this creates some connection between envy-freeness and our framework, we will explain in Section \ref{ssec:Limitations} why group-level envy-freeness does not fall neatly into our framework.

Most of the discussed works so far have taken a parity-based approach, which \cite{binns2020apparent, kuppler2021distributive} connect to (luck) egalitarianism. 
\cite{hu2020fair} already mentions that one may want to maximize the utility of the marginalized group to overcome the ``leveling down" objection against parity-based metrics.
Indeed, the distributive justice literature offers many more distribution patterns than egalitarianism \cite{kuppler2021distributive}.
Some expansions of group fairness have looked at these other patterns.
\cite{martinez2020minimax} and \cite{diana2021minimax} do not attempt to equalize harm across groups but to minimize the harm of the group with the highest error rate (referred to as minimax). The diametrically opposed maximin principle, which was popularized by John Rawls~\cite{rawls1999theory, rawls2001justice}, maximizes the benefit or utility of the worst-off group. \cite{Franke2021} describes the maximin principle as being useful in high-stakes decision-making where risk aversion is appropriate.
These works take important steps in considering other distributions of utility than parity-based ones. However, they still only look at one specific pattern and do not discuss how this fits into a general framework of group fairness.

So while several works expand on standard group fairness criteria, none of them provides a comprehensive framework that integrates different theories of distributive justice.
The goal of this paper is to propose such a framework.

\section{A framework for fairness evaluations based on distributive justice}\label{sec:framework}

We consider a decision-making system that takes binary decisions $D$ on decision subjects (DS) of a given population $P$, based on a decision rule $r$.
The decision rule assigns each individual $i \in P$ a binary decision $d_i \in \{0,1\}$, which depends on an unknown but decision-relevant binary random variable $Y$, by applying $r$ to some input data.
The decision rule could, for example, be an automated rule that takes decisions based on predictions of $Y$ via a predicted score $S \in [0,1]$ for every individual, derived from an ML model,  or the decisions could be made by humans. We assume that at least two socially salient groups are defined, denoted by different values $a$ for the sensitive attribute $A$.
In the following, we first introduce the four components of our framework for group fairness. 
Then, in Section~\ref{ssec:general_definition}, we provide a generalized definition of group fairness, which encompasses all components of the proposed framework.

\subsection{Modeling consequences: Utility of the decision subjects}\label{sec:utility-weights}

For modeling the consequences of decisions for decision subjects, we use a utility function $u_{DS}$ which, in our binary context, may depend on both the decision $d_i$ and the value $y_i$ of $Y$. $u_{DS}$ is positive in the case of a benefit, and negative in the case of a harm. 
In the simplest case, we ignore individual differences in the utility function. Then, the utility $u_{DS,i}$ of a decision subject $i$ with $Y=y_i$ subjected to a decision $D=d_i$ is given by:
\begin{equation}\label{eq:utility-ds-individual}
u_{DS,i} = w_{11} \cdot d_i \cdot y_i + w_{10} \cdot d_i \cdot (1-y_i) + w_{01} \cdot (1-d_i) \cdot y_i + w_{00} \cdot (1-d_i) \cdot (1-y_i),
\end{equation}
where $w_{dy}$ denote the four different utility values that might be realized for the four combinations of the random variables $Y$ and $D$, leading to the utility matrix $W=(w_{00},w_{01},w_{10},w_{11})$.%
\footnote{
More complex utility functions can be used, up to a fully individualized utility function. A simple extension would also take $A$ into account and define the utility matrix for each group separately, i.e., using utility weights of all possible outcomes that depend on the group membership ($w_{dy} \not\perp a$).
In philosophy and economics, the work of Amartya Sen explains why resources do not always convert into the same capabilities (options to be and do) \cite[pp. 21-23]{sen1985standard}, which would suggest such an extension.
}
The utility $u_{DS,i}$ is a realization of a random variable $U_{DS}$. For assessing the fairness of a decision-making system, we are interested in \emph{systematic} differences between groups. We follow the standard group fairness assumption that such differences correspond to different expectation values $E(U_{DS})$ for different groups in $A$~\cite{fairmlbook}.

\subsection{Defining groups: Relevant groups}\label{sec:relevant-groups}

Group fairness is concerned with socially salient groups (e.g., defined by gender or race) as this is what theories of discrimination focus on \cite{sep-discrimination}.
We refer to these groups as \textit{relevant groups}, denoted by $A$, and expect them to at least have a \textit{weak causal influence} on the prediction or outcome (or both). 
This means we can plausibly expect group membership in the relevant groups to be a (direct or indirect) cause of inequalities.
In~\cite{Hertweck2023JusticeBasedFramework}, a philosophical argument is provided for this definition. 

\subsection{Defining subgroups: Claim differentiator}
\label{ssec:Claims_differentiator}

Comparing the relevant groups as such might not always be morally appropriate. For example, equality of opportunity \cite{hardt2016equality} only considers individuals with $Y=1$. This might be considered morally appropriate if the moral claim for a positive decision depends on $y_i$. In our framework, we allow for a so-called \textit{claim differentiator}, represented by a feature $J$ which differentiates individuals with different claims to the utility. Different claims may be justified, e.g., by differences in deservingness, need, or merit. All individuals with the same value $J=j$ are considered to have the same claim to utility.
\footnote{A similar idea is found in~\cite{loi2021fair,baumann2022SDS_fairness_principle,holm2022fairness}.} 
Consequently, comparing relevant groups $a$ may be conditioned on subgroups with an equal moral claim (hence equal value $j$): Instead of $E(U_{DS}|A=a)$, we compare $E(U_{DS}|J=j, A=a)$. Note that not all possible values $j$ might be considered  relevant from a fairness perspective. 

\subsection{Just distribution: Pattern of justice}\label{sec:patterns}

The claim differentiator $J$ defines which individuals have equal moral claims to the utility distributed by the decision process. One might argue that this calls for equal utility. However, the literature on distributive justice shows that this is not necessarily the case.
Our approach thus offers additional normative choices, which we refer to as \textit{patterns of justice}. For each of them, we will briefly explain their normative view of what constitutes justice and formulate a \textit{fairness constraint}, representing a mathematical formalization of a pattern of justice, which can either be satisfied or not.
For simplicity, we restrict ourselves to the case of two relevant groups $A=\{0, 1\}$ even though our framework generalizes to more groups. 

In the following, we introduce only a few patterns of justice (representing fairness principles for the distribution of benefits) that are widely discussed in the philosophical literature.
However, our utility-based definition of group fairness should in no way be seen as limited to these patterns.

\subsubsection{Egalitarianism}

Egalitarianism -- as the name suggests -- demands equality \cite{sep-egalitarianism}.
However, egalitarianism as a broad concept does not specify \textit{what} should be equalized. This is the subject of the \textit{equality of what} debate initiated by \cite{sen1980equality}. One could, e.g., aim to equalize the opportunities (equality of opportunity) or outcomes (equality of outcomes). In our approach, we consider utility as the quantity that has to be equalized. 

\paragraph{Fairness criterion} The egalitarian fairness criterion is satisfied if the expected utility is equal for the relevant groups conditioned on the claim differentiator:
\begin{equation}\label{eq:egalitarianism-goal}
E(U_{DS}|J=j, A=0)=E(U_{DS}|J=j, A=1)
\end{equation}

\subsubsection{Maximin}

Maximin describes the principle that among a set of possible distributions, the one that maximizes the expected utility of the group that is worst-off should be chosen \cite{sep-social-choice}. In contrast to egalitarianism, inequalities are thus tolerated if the worst-off group benefits from them. This has been defended by Rawls in the form of the ``difference principle'' \cite{rawls1999theory, rawls2001justice}.

\paragraph{Fairness criterion} A decision rule $r'$ satisfies the maximin fairness criterion if there is no other possible rule $r$ that would lead to a greater expected utility of the worst-off group $E(U_{DS})^{worst-off}$. 
\begin{equation}\label{eq:maximin-goal}
E(U_{DS})^{worst-off}(r') \geq max_{r \in R} \left( E(U_{DS})^{worst-off}(r) \right)
\end{equation}

\subsubsection{Prioritarianism}

Prioritarianism describes the principle that among a set of possible distributions, the one that maximizes the weighted sum of utilities across all people should be chosen, where the utility of the worst-off group is given a higher weight \cite{holtug2017prioritarianism}. Thus, the normative goal is to maximize $\tilde U_{DS} = k \cdot E(U_{DS})^{worst-off} + E(U_{DS})^{better-off}$, with $k>1$.  
\footnote{
The maximin principle can be seen as the extreme version of this as an infinite weight is given to the worst-off relevant groups).}

\paragraph{Fairness criterion} A decision rule $r'$ satisfies the prioritarian fairness criterion  if
\begin{equation}\label{eq:prioritarianism-goal}
\tilde U_{DS}(r') \geq max_{r \in R} \left( \tilde U_{DS}(r) \right),
\end{equation}

\subsubsection{Sufficientarianism}

Sufficientarianism \cite{shields2020sufficientarianism} describes the principle that there is a minimum threshold of utility that should be reached by everyone in expectation. Inequalities between relevant groups are acceptable as long as all groups achieve a minimum level of utility in expectation.

\paragraph{Fairness criterion} The sufficientarian fairness criterion is satisfied if all groups' expected utilities are above a given threshold $t$:
\begin{equation}\label{eq:sufficientarianism-goal}
\forall a \in A \;\; E(U_{DS}|J=j, A=a) \geq t
\end{equation}

\subsection{Generalized definition of group fairness}
\label{ssec:general_definition}

Instead of seeing group fairness as demanding equality between socio-demographic groups with respect to a statistical quantity, we propose the following generalized definition:

\begin{definition}[Group fairness]
\emph{Group fairness} is the just distribution of utility among groups, as defined by the specification of a utility function, relevant groups, a claim differentiator, and a pattern of justice. \textit{Group fairness criteria} specify when group fairness is satisfied by a decision-making system.
\end{definition}

We will show that the standard group fairness criteria are special cases of this definition of group fairness with different utility functions and claim differentiators. However, all of them are based on the pattern of egalitarianism. The proposed generalization allows for arbitrary utility matrices, yielding the possibility to compare consequences rather than decisions, and additional patterns of justice, as suggested by the relevant philosophical literature.

This extension of group fairness criteria alleviates some of the criticisms of currently popular group fairness criteria as we will show in Section \ref{sec:discussion}.

\section{Relation to standard group fairness criteria}
\label{sec:relation-group-fairness}

Standard group fairness criteria derived from the confusion matrix are special cases of the group fairness framework that we propose.
They follow the egalitarian pattern of justice and correspond to specific decision subject utility functions ($U_{DS}$), and specific choices for the claim differentiator $J$ and its considered values $j$. Table~\ref{tab:utility-matrix-representation} shows the mapping of our framework to standard group fairness criteria.
For example, the acceptance rate is equivalent to the expected DS utility ($E(U_{DS})$) without any claim differentiator ($J=\emptyset$) if the utility weights are chosen as $w_{11}=w_{10}=1$ and $w_{01}=w_{00}=0$.
Similarly, for $J=Y$, $j \in \{1\}$, and $w_{11}=w_{01}=1$, the expected DS utility ($E(U_{DS}|Y=1)$) corresponds to the true positive rate.

\begin{table*}
\small
\centering
\caption{Utility matrix representation of metrics used for standard group fairness criteria. Gray patches depict unused DS utility weights due to the claim differentiator $J=j$. The DS utility weights ($w_{dy}$) are represented as \parbox[c]{2em}{\includegraphics[width=0.4in]{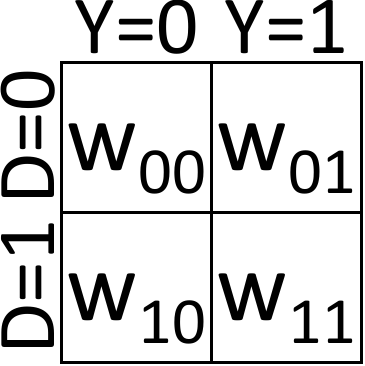}}
}
\label{tab:utility-matrix-representation}
\begin{tabular}{llll} 
\toprule
\textbf{$U_{DS}$}                                  & \textbf{$J$}           & $j$       & \textbf{Metric}                  \\
\midrule
\parbox[c]{1em}{\includegraphics[width=0.25in]{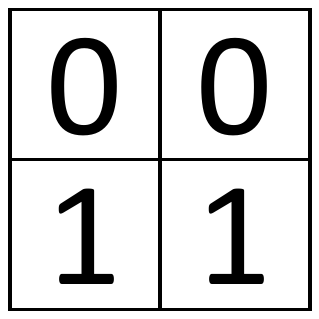}}           & $\emptyset$   & -         & Acceptance rate                    \\ \hline
\parbox[c]{1em}{\includegraphics[width=0.25in]{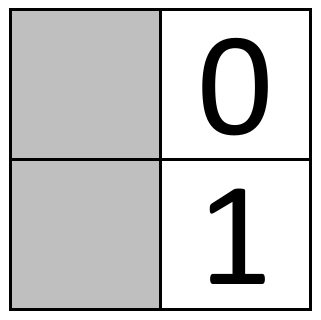}} & $Y$           & $\{1\}$   & True positive rate               \\ \hline
\parbox[c]{1em}{\includegraphics[width=0.25in]{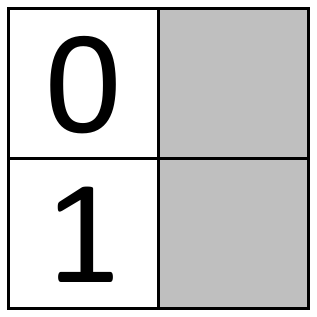}} & $Y$           & $\{0\}$   & False positive rate            \\ \hline
\parbox[c]{1em}{\includegraphics[width=0.25in]{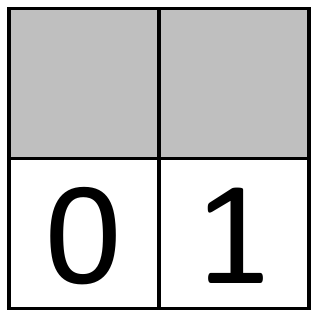}} & $D$           & $\{1\}$   & Positive predictive value                     \\ \hline
\parbox[c]{1em}{\includegraphics[width=0.25in]{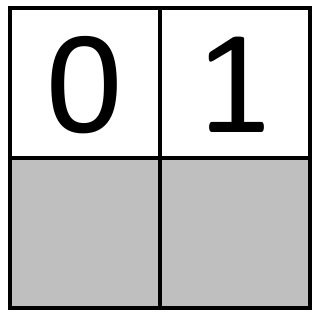}} & $D$           & $\{0\}$   & False omission rate            \\
\bottomrule
\end{tabular}
\end{table*}

\subsection{Standard group fairness criteria through the lens of our utility-based approach}
\label{ssec:proofs}

The examples in Table~\ref{tab:utility-matrix-representation} are not the only possibilities of utility matrices that lead to equivalence with a standard group fairness metric. Equality of $U_{DS}$ between two relevant groups is insensitive against some changes of the utility matrix $W$. In particular, adding a constant to all matrix elements, or multiplying them with a constant factor, does not change the fairness criterion.%
\footnote{This allows for choosing a convenient reference point for utility, e.g. setting one of the elements to $0$. By defining another one to have the value of $1$, a scaling is introduced. See also~\cite{elkan2001} for a discussion of this topic.}
Thus, different utility matrices may lead to an equivalence to one of the standard group fairness criteria. In this section, we show under which conditions we achieve such an equivalence 
(see Table~\ref{tab:relation-to-group-fairness} for a summary of the results w.r.t.: (conditional) statistical parity,\footnote{Notice that the claim differentiator for conditional statistical parity is defined as $J=L$, where $L$ denotes legitimate attributes that can take the values $l$.} equality of opportunity, false positive rate (FPR) parity, equalized odds, predictive parity, false omission rate (FOR) parity, and sufficiency.\footnote{Note that we focus on fairness criteria that are based on the decisions $D$ and actual outcomes $Y$. However, this idea generalizes to fairness definitions that are based on predicted scores and actual outcomes, such as balance for the positive/negative class and well-calibration (see ~\cite{kleinberg2016inherent, chouldechova2017fair, verma2018fairness} for a definition of these criteria).}
The mathematical definitions of these criteria can be found in Table~\ref{tab:group-fairness-criteria} in Appendix~\ref{sec:appendix-group-fairness}.
In the following, we focus on statistical parity, equality of opportunity, and predictive parity as prototypical examples.
We refer the interested reader to the Appendix~\ref{appendix:additional_results} for a similar mapping of other standard group fairness criteria.

\begin{table*}
\small
\centering
\caption{Mapping of standard group fairness metrics to our utility-based approach under Egalitarianism}
\label{tab:relation-to-group-fairness}
\begin{tabular}{llll}
\toprule
\multicolumn{3}{l}{\textbf{General conditions}}                                                               & \textbf{Equivalent fairness criterion}\\
$U_{DS}$ weights (for groups $a \in \{0,1\}$)                                 & $J$           & $j$       & \multicolumn{1}{l}{}                  \\
\midrule
$w_{11} = w_{10} \neq w_{01} = w_{00} \; \land \; w_{dy} \perp a$           & $\emptyset$   & -         & Statistical parity                    \\ \hline
$w_{11} = w_{10} \neq w_{01} = w_{00} \; \land \; w_{dy} \perp a$           & $L$           & $l$       & Conditional statistical parity        \\ \hline
$w_{11} \neq w_{01} \; \land \; w_{d1} \perp a$                             & $Y$           & $\{1\}$   & Equality of opportunity               \\ \hline
$w_{10} \neq w_{00} \; \land \; w_{d0} \perp a$                             & $Y$           & $\{0\}$   & False positive rate parity            \\ \hline
$w_{11} \neq w_{01} \land w_{10} \neq w_{00} \; \land \; w_{dy} \perp a$    & $Y$           & $\{0,1\}$ & Equalized odds                        \\ \hline
$w_{11} \neq w_{10} \; \land \; w_{1y} \perp a$                             & $D$           & $\{1\}$   & Predictive parity                     \\ \hline
$w_{01} \neq w_{00} \; \land \; w_{0y} \perp a$                             & $D$           & $\{0\}$   & False omission rate parity            \\ \hline
$w_{11} \neq w_{10} \land w_{01} \neq w_{00} \; \land \; w_{dy} \perp a$    & $D$           & $\{0,1\}$ & Sufficiency                           \\
\bottomrule
\end{tabular}
\end{table*}

\textbf{Statistical parity} (also called demographic parity or group fairness~\cite{dwork2012fairness}) is defined as $P(D=1|A=0)=P(D=1|A=1)$.
\begin{proposition}[Statistical parity as utility-based fairness]
    If the utility weights of all possible outcomes (as defined in Section~\ref{sec:utility-weights}) do not depend on the group membership ($w_{dy} \perp a$), and $w_{11} = w_{10} \neq w_{01} = w_{00}$, then the egalitarian pattern fairness condition with $J=\emptyset$ is equivalent to statistical parity.
\label{prop:statistical_parity_as_utility}
\end{proposition}
The formal proof of Proposition~\ref{prop:statistical_parity_as_utility} can be found in Appendix~\ref{proof:statistical_parity}.

To measure the different degrees to which egalitarian fairness is fulfilled, we can introduce a quantitative fairness \textit{metric} $F$. One option is to compute the absolute difference between the two groups' expected utilities:%
\footnote{
Note that other metrics could be used, e.g., the ratio of the two expected utilities.
}
\begin{equation}\label{eq:egalitarianism-measurement}
F_{\text{egalitarianism}} = |E(U_{DS}|J=j, A=0) - E(U_{DS}|J=j, A=1)|
\end{equation}
Up to a multiplicative constant, this measure is equivalent to the degree to which statistical parity is fulfilled:
\begin{corollary}[Partial fulfillment of statistical parity in terms of utility-based fairness]
    Suppose that the degree to which statistical parity is fulfilled is defined as the absolute difference in decision ratios across groups, i.e., $|P(D=1|A=0) - P(D=1|A=1)|$.
    If the utility weights do not depend on the group membership ($w_{dy} \perp a$), and $w_{11} = w_{10} \neq w_{01} = w_{00}$ (i.e., $w_{1y} \neq w_{0y}$), and $J=\emptyset$, then the degree to which egalitarianism is fulfilled is equivalent to the degree to which statistical parity is fulfilled, multiplied by $|w_{1y} - w_{0y}|$.
    \label{cor:partial_fulfillment_of_statistical_parity}
\end{corollary}
The formal proof of Corollary~\ref{cor:partial_fulfillment_of_statistical_parity} can be found in Appendix~\ref{proof:partial_fulfillment_of_statistical_parity}.

\textbf{Equality of opportunity} (also called TPR parity) is defined as $P(D=1|Y=1, A=0)=P(D=1|Y=1, A=1)$, i.e., it requires parity of true positive rates (TPR) across groups $a \in A$~\cite{hardt2016equality}.
In this case, not all values of the claim differentiator $J$ are considered to be relevant: we are only concerned with individuals of type $Y=1$.
\begin{proposition}[Equality of opportunity as utility-based fairness]
    If $w_{11}$ and $w_{01}$ do not depend on the group membership ($w_{d1} \perp a$), and $w_{11} \neq w_{01}$, then the egalitarian pattern fairness condition with $J=Y$ and $j \in \{1\}$ is equivalent to equality of opportunity.
\label{prop:EOP_as_utility}
\end{proposition}
The formal proof of Proposition~\ref{prop:EOP_as_utility} can be found in Appendix~\ref{proof:EOP_as_utility}.
Compared to statistical parity, equality of opportunity only requires equal acceptance rates across those subgroups of $A$ who are of type $Y=1$.
This corresponds to the claim differentiator $J=Y$ with $j \in \{1\}$.
\footnote{See Corollary~\ref{cor:partial_fulfillment_of_equality_of_opportunity} in Appendix~\ref{app:Additional_corollaries} for the extension to a partial fulfillment of equality of opportunity.}

\textbf{Predictive parity} (also called PPV parity~\cite{baumann2022sufficiency} or outcome test~\cite{simoiu2017problem}) is defined as $P(Y=1|D=1, A=0)=P(Y=1|D=1, A=1)$. It requires parity of positive predictive value (PPV) rates across groups $a \in A$.
\begin{proposition}[Predictive parity as utility-based fairness]
    If $w_{11}$ and $w_{10}$ do not depend on the group membership ($w_{1y} \perp a$), and $w_{11} \neq w_{10}$, then the egalitarian pattern fairness condition with $J=D$ and $j \in \{1\}$ is equivalent to predictive parity.
    \label{prop:predictive_parity_as_utility}
\end{proposition}
The formal proof of Proposition~\ref{prop:predictive_parity_as_utility} can be found in Appendix~\ref{proof:predictive_parity_as_utility}.
\footnote{See Corollary~\ref{cor:partial_fulfillment_of_predictive_parity} in Appendix~\ref{app:Additional_corollaries} for the extension to a partial fulfillment of predictive parity.}%
\textsuperscript{,}
\footnote{
Notice that the fairness notion \textit{well-calibration} is related to PPV parity but it is defined for scores instead of binary decisions: $P(Y = 1|S = s,A=0) = P(Y = 1|S = s,A=1)$. This requires that for each predicted score $s \in S$, individuals of all groups $a$ have equal chances of belonging to the positive class~\cite{chouldechova2017fair}.
Our proposed approach is equivalent to satisfying well-calibration if $J=S$, $w_{s1}=1$, $w_{s0}=0$, and using an egalitarian pattern of justice. 
In this case, the claim differentiator is the predicted score $s$, and all possible values of $s$ need to be considered.
Notice that, in this special case, the DS utility weights (denoted by $w_{sy}$) only depend on $Y$ and are uniform across the entire range of scores. If one wants to extend well-calibration to take score-specific consequences of outcomes into account, this can be done easily by introducing score-specific utilities $w_{sy}$.
The stronger definition of well-calibration ($P(Y = 1|S = s,A=0) = P(Y = 1|S = s,A=1) = s$), which is sometimes also called \textit{calibration by groups}~\cite{fairmlbook} or \textit{calibration within groups}~\cite{kleinberg2016inherent,verma2018fairness}, is equivalent to requiring that $\forall a \in A, \forall s \in S \;\;E(U_{DS}|J=s,A=a)=s$, for $J=S$, $w_{s1}=1$, $w_{s0}=0$. Here, the pattern is stronger than just egalitarianism.
}

\subsection{Uncovering the moral assumptions of standard group fairness metrics}

Considering Table~\ref{tab:relation-to-group-fairness}, we see that each standard group fairness criterion (a) constitutes a specific way of measuring the benefit/harm of decision subjects, (b) embeds assumptions about who has equal or different moral claims to utility, and (c)
requires equality. All these elements correspond to normative choices that define what kind of fairness is achieved.

If we were to, for example, demand equality of opportunity for men and women in credit lending (where $D$ is the bank's decision to either approve a loan ($D=1$) or reject it ($D=0$), and $Y$ is the loan applicant's ability to repay the loan ($Y=1$) or not ($Y=0$)), we make the following assumptions: The benefit derived by being granted a loan is the same for each individual and the same for men and women. Only people who repay their loans have a legitimate claim to utility, and we don't need to consider the consequences for people who do not repay. Fairness means equalizing the acceptance rates of men and women of the morally relevant group (those who would repay a granted loan), even if this leads to undesirable outcomes for both men and women -- other solutions are not considered. 

All of these assumptions can be disputed for good reasons. For example, should we really ignore that being granted a loan might not only be beneficial for someone who cannot repay it? And is it morally acceptable to ignore the consequences for the defaulters? Also, is it really desirable to make every group worse off just for the sake of equality?  These questions come up naturally when we analyze the utility matrix, the relevant groups, the claim differentiator, and the pattern of justice. Our framework shows possible alternatives for each component. This helps considerably to decide whether or not the chosen fairness criterion is morally appropriate and forces stakeholders to make their moral assumptions explicit, which are usually left implicit in standard approaches for choosing between group fairness criteria.

\section{A simple application example}\label{sec:example}

Suppose that an ML-based decision-making system is used to identify those patients in a cancer population that will benefit from an innovative drug. 
Patients from the positive class ($Y=1$) do not develop side effects after the drug treatment (or the side effects are negligible), i.e., they would benefit from the treatment because it cures their cancer.
But those from the negative class ($Y=0$) suffer from side effects of the cancer treatment.
For the sake of this argument, let us assume that, despite being cured of cancer, those side effects require another treatment, which reduces life quality significantly over the next year. 
Due to the high cost of both treatments (the one against cancer and the one to treat the potential side effects), only individuals with a high likelihood of not developing any side effects ($p$) are treated ($D=1$).
More specifically, we assume that the optimal decision from the perspective of the decision makers (e.g., the hospital) would be to treat all individuals with a probability $p$ that lies above 50\% $\left(p=P\left(Y=1\right)>0.5\right)$.
We further assume that due to the non-representative selection of the research subjects for clinical trials, individuals from the minority group are much more likely to suffer from side effects (i.e., have lower probabilities $p$ of not developing side effects). Absent any fairness considerations, this results in a lower treatment rate for the minority group.
One might argue that the selection of cancer patients for treatment with the new drug should be made in a fair manner to avoid disadvantaging individuals in the minority group. 
This requires the elicitation of a morally appropriate group fairness metric.
First, we will use established methods to select a standard metric.
Then, we will apply our proposed utility-based approach. We will compare the results of both methods and analyze their implications.

First, using existing approaches to select one of the standard group fairness criteria~\cite{heidari2019moral,loi2021fair,baumann2022SDS_fairness_principle}, one might argue that statistical parity is an appropriate choice because the likelihood of requiring additional expensive treatment (due to developing side effects) does not determine how deserving people are to live without cancer -- even if this may be a relevant consideration for efficiency reasons absent any fairness constraints. Thus, the chances of treatment should be equal for individuals of both groups. 

Second, we elicit a morally appropriate fairness criterion by going through the four components of our framework:
Regarding the definition of the relevant groups to compare, the example's assumption is that it is the minority group for which fairness should be ensured in comparison to the majority group.
For example, it may be argued that this is reasonable due to the causal link between a patient's group membership and the likelihood of developing side effects (see Section~\ref{sec:relevant-groups}).
As for the claim differentiator, it may be assumed that all individuals have the same moral claim to utility, i.e., that there is no justifiable argument to differentiate between individuals' deservingness (or necessity or urgency) to be treated.
In this case, following the same moral standpoint as above, there would not be any claim differentiator ($J=\emptyset$), equivalent to the case of statistical parity.
However, critical differences may emerge when going through the other two steps of our framework, i.e., the evaluation of the utility of the DS (as introduced in Section~\ref{sec:utility-weights}) and the specification of the pattern for a just distribution of the utility derived by the cancer patients (see Section~\ref{sec:patterns}).

\begin{figure}[tb]
\centering
\begin{subfigure}[b]{0.4\textwidth}
\centering
\includegraphics[width=0.4\textwidth]{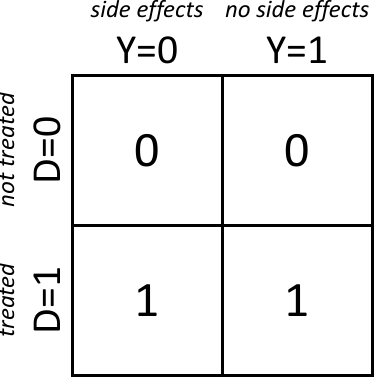}
\caption{Statistical parity}
\label{fig:confusion_matrix_health_1}
\end{subfigure}
\begin{subfigure}[b]{0.4\textwidth}
\centering
\includegraphics[width=0.4\textwidth]{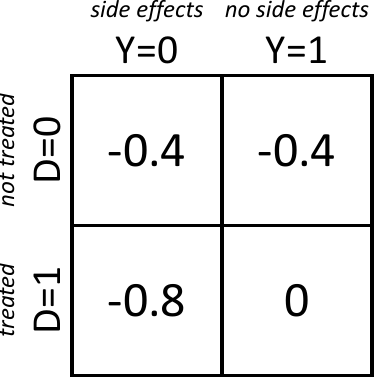}
\caption{DS utility with considerable side effects}
\label{fig:confusion_matrix_health_3}
\end{subfigure}
\caption{
Decision subject utilities for the medical example. Left, the DS utility matrix for the standard group fairness criteria statistical parity is shown.
The matrix on the right represents the DS utility (in DALYs, here represented with negative utility) for the medical drug treatment example with considerable, burdensome side effects for the treated patients.}
\label{fig:confusion_matrix_health}
\end{figure}

Let us now specify the DS utility.
Using the disability-adjusted life years (DALY)\footnote{
DALY is a generic measure of disease burden calculated as the sum of the years of life lost (YLL) due to dying early and the years lost due to disability or disease (YLD), i.e., DALY = YLL + YLD, where one DALY represents the loss of the equivalent of one year of full health.
} as a measure for patients' negative expected utilities to compare different outcomes of the medical treatment, we may specify the DS utility as follows:
Individuals without any side effects receiving the treatment can live a cancer-free life, defining our reference point: $w_{11}=0$ (representing zero DALYs).
Individuals that do not receive the treatment continue living for one more year with the disease burden: $w_{00}=w_{01}=-0.4$ (representing slightly less than half a DALY).
The utility of individuals developing side effects after having received the treatment depends on the severity of those side effects.
For this simple example, we assume that the side effects are considerable but do not result in death.
More precisely, we assume that the burden of the side effects and the additional treatment is equivalent to -0.8 DALYs (i.e., for the assumed year of life, $w_{10}=-0.8$).
This is represented by the utility matrix in Fig.~\ref{fig:confusion_matrix_health_3}, next to the DS utility matrix equivalent to the standard group fairness criteria statistical parity in Fig.~\ref{fig:confusion_matrix_health_1}.

Next, we need to specify a pattern of justice, which defines what a just distribution looks like.
We assume that maximizing the expected utility of the worst-off group (i.e., a maximin DS utility distribution) is desirable from a fairness perspective, as one might reasonably argue in a risk-averse health context~\cite{Franke2021}, which would be in line with Rawls' initial original position~\cite{rawls1999theory}.

This example shows that our general framework results in a different fairness metric. Not only is the benefit measured differently because we are taking the consequences of a decision (including possible side effects) into account, but we also apply a different pattern of justice. 

We will now show that enforcing statistical parity does not necessarily make the minority group better off, on average.
To ensure statistical parity, more individuals from the minority group have to be treated, compared to the unconstrained optimum, since minority individuals have a systematically lower $p$.
However, whether being treated (i.e., switching from $D=0$ to $D=1$) is desirable for the patients, depends on the side effects: those who do not develop side effects gain utility (i.e., their expected utility changes from -0.4 to 0) and those who do develop side effects lose utility (i.e., their expected utility changes from -0.4 to -0.8).
Apart from degenerate cases ($p=0$ and $p=1$), patients do not know with certainty if they will develop side effects, as the outcome $Y$ is unknown.
In expectation, a treatment is only desirable for individuals with $p>= \frac{w_{00} -w_{10}}{w_{11} - w_{01} - w_{10} + w_{00}} = 0.5$.
Hence, increasing the number of treated minority patients is problematic, as the patients of the minority group who are treated additionally experience a disadvantage by the treatment rather than an advantage\footnote{Recall that, without fairness consideration, the hospital decided to treat patients with $p>0.5$, due to cost considerations. Thus, increasing the number of treated patients in a group requires treating patients with $p<0.5$}. 
This is completely disregarded by the fairness metric statistical parity, which implicitly assumes that a positive decision is desirable for anyone ($w_{11} = w_{10} =1$ and $w_{01} = w_{00}=0$).
In fact, in this scenario, enforcing statistical parity would likely make both groups worse off (by increasing/decreasing the number of treated patients in the minority/majority group), compared to the unconstrained case, in order to equalize the share of treated patients in the two groups -- leading to a classical case of the ``leveling down objection''.

Applying the maximin pattern of justice, in contrast, can prevent us from producing `fairness' at the cost of the minority group, which would contradict the overall goal of improving the situation for the minority group.

\section{Discussion}\label{sec:discussion}

In this section, we discuss how our proposed framework alleviates the previously discussed limitations of standard group fairness criteria, and we comment on the limitations of our expanded definition of group fairness.

\subsection{Alleviating limitations of existing fairness criteria}

Standard group fairness criteria are special cases of our generalized group fairness framework. The suggested extension allows the alleviation of several of the standard group fairness limitations that we discussed in Section \ref{sec:limitation-group-fairness}.

\paragraph{The ``leveling down objection''}

The ``leveling down objection'' is a prevalent anti-egalitarianism argument \cite{parfit1995,crisp2003} saying that less inequality is not desirable if this requires lowering the better-off group's utility to match the one of the worse-off group.
On this basis, choosing egalitarianism as the pattern of justice has been criticized in the algorithmic fairness literature (see, e.g., \cite{loi2021fair,hu2020fair,weerts2022does}). Our approach allows using other patterns of justice, such as maximin, prioritarianism, or sufficientarianism (see Section \ref{sec:patterns}). Other patterns that can be formalized as mathematical constraints may also be used. One could, for example, combine several patterns into one and require equal expected utilities across groups as long as none of the groups is better off than it would be without any fairness requirement. This would represent a combination of egalitarianism and a group-specific baseline threshold (similar to sufficientarianism), making a ``leveling down'' of the better-off group impossible and adhering to the Pareto principle.

\paragraph{Focus on decisions instead of consequences}

Standard group fairness criteria only consider the distribution of \textit{either} $D$ \textit{or} $Y$. This can be interpreted as analyzing the distribution of utility but assuming that utility is equivalent to \textit{either} $D$ \textit{or} $Y$ instead of, for example, the combination of $D$ \textit{and} $Y$. Standard group fairness criteria thus represent a very confining definition of utility. Our approach acknowledges that the utility of the decision subjects may depend on a combination of different attributes such as one's ability to repay a loan or one's socioeconomic status (see, e.g., \cite{hertweck2021moral,weerts2022does,binns2018fairness}. This is represented through the utility function described in Section \ref{sec:utility-weights}, which can easily be extended (e.g., to take group-specific utility functions into account).

\paragraph{Limited set of fairness definitions}

Previous attempts to guide stakeholders in choosing appropriate fairness criteria have taken on the form of explicit rules, such as in \cite{saleiro2018aequitas,makhlouf2021applicability,ruf2022tool}. 
Works like~\cite{heidari2019moral,loi2021fair,baumann2022SDS_fairness_principle} have provided unifying moral frameworks for understanding existing notions of algorithmic fairness, but they still presuppose a limited set of fairness definitions from which stakeholders can choose.
While~\cite{heidari2019moral} consider the distribution of \textit{undeserved} utility (what they call the \textit{difference between an individual's actual and effort-based utility}), \cite{loi2021fair} and~\cite{baumann2022SDS_fairness_principle} use the decision subject utility $U_{DS}$ to derive a morally appropriate group fairness definition.
This is similar to the approach presented in this paper; however, they only consider two options $U_{DS}=D$ and $U_{DS}=Y$, while our approach allows for arbitrary functions $f$ for the utility: $U_{DS}=f(D,Y)$.
Furthermore, \cite{heidari2019moral,loi2021fair,baumann2022SDS_fairness_principle} only consider egalitarian notions of fairness, and it remains unclear how non-egalitarian notions of fairness fit in.

As discussed in Section \ref{sec:extension}, previous works have expanded standard group fairness metrics. However, the resulting fairness notions diverge from standard metrics on some of the four components in our framework. The other components are held constant compared to standard group fairness metrics, while the assumptions that are encompassed in the choice to keep these components constant are not made explicit. Therefore, the criteria resulting from these expansions are still somewhat limited.
We provide a method that integrates these prior works in a unifying framework and link the different choices to morally relevant concepts with respect to the utility function for decision subjects (Section \ref{sec:utility-weights}), the relevant groups to compare (Section \ref{sec:relevant-groups}), the subgroups with equal claims to utility (Section \ref{ssec:Claims_differentiator}), and the pattern for a just distribution of utility (Section \ref{sec:patterns}).

\subsection{Limitations}
\label{ssec:Limitations}

\paragraph{Fundamental assumptions of standard group fairness criteria}

While our framework extends standard group fairness criteria, we still share some of the fundamental assumptions embedded in group fairness.
First, we compare averages across groups, which has been criticized for being vulnerable to fairness gerrymandering \cite{dwork2012fairness}. There could be systematic differences between groups despite them having the same averages, e.g., due to a different distribution within groups.
Second, one could criticize that we (and group fairness notions in general) cannot distinguish cases in which membership in the relevant groups has a causal influence on the outcomes and decisions or whether they just happen to be correlated -- in cases where both result in the same distribution of utilities. Contrary to that, counterfactual fairness \cite{kusner2017counterfactual} demands that the sensitive attribute (and its proxies) do not influence the final decision.
While we cannot guarantee that our group fairness criteria would fulfill such a strict requirement, we argue that our approach to group fairness most likely avoids the objection of fairness gerrymandering and causal irrelevance in practice.
Our practical solution to these objections is to require at least a weak causal link for the specification of relevant groups (as mentioned in Section \ref{sec:relevant-groups}): We demand that individuals belong to a relevant group that is likely to be the cause of an unjust inequality. 
This way, we reduce the probability that group fairness is evaluated in a situation where the inequality is caused by spurious unfortunate correlations.
When defining the relevant groups, we could make them increasingly narrow. This idea of increasingly narrow groups aligns with the concept of multicalibration, which is motivated by the concept of individual fairness.
It can be seen as a further extension of well-calibration. Multicalibration calls for calibrating every efficiently-identifiable subgroup $R$ of a computationally-identifiable subset $\mathcal{C}$ of the population $P$~\cite{hebert-johnson18Multicalibration}. Intuitively, multicalibration can also be seen as a special case of our proposed framework, where $\forall r \in R \;\;E(U_{DS}|J=s,R=r)=s$ for all subgroups $R \in \mathcal{C}$, for $J=S$, $w_{s1}=1$, $w_{s0}=0$.
However, the larger the number of subgroups, the more difficult it becomes to make moral judgments about them. Therefore, instead of only considering subgroups based on computational efficiency as in multicalibration, we focus on groups that meet the weak causality requirement.
Furthermore, w.r.t. multicalibration, our proposed framework demonstrates that benefits/harms are measured in a narrow way ($w_{s1}=1$, $w_{s0}=0$), which can be extended using the flexibility of the DS utility function.
Our framework thus again reveals the normative choices of these fairness notions.

\paragraph{Economic notions of fairness}

As we explained in Section \ref{sec:extension}, our framework builds on existing extensions of standard group fairness metrics and tries to structure these. Yet, there are still some extensions that do not fit neatly into our framework. As far as we are aware, this mainly concerns \citet{zafar2017FromParitytoPreference}'s interpretation of group-level envy-freeness for fair machine learning. Contrary to \citet{hossain2020designing}, they postulate that group-level envy-freeness is fulfilled if ``every sensitive attribute group (e.g., men and women) prefers the set of decisions they receive over the set of decisions they would have received had they collectively presented themselves to the system as members of a different sensitive group'' \cite[p.~3]{zafar2017FromParitytoPreference}. This fairness criterion is structurally different from the fairness criteria in our framework: Our fairness criteria compare the average utilities of different groups. Instead, the envy-freeness criterion compares the average utility of a single group to the expected utility of this group if this group had a different sensitive attribute -- it thus compares the average utility of a single group under different assumptions. Similarly, \citet{kim2019PreferenceInformedFairness}'s preference-informed statistical parity compares utilities of groups across alternative classifiers instead of comparing utilities between groups for a single classifier. The question is thus not about how a classifier should distribute utilities between equally deserving groups but about whether a classifier makes every group better off than some alternative.

\paragraph{Theories of distributive justice}
While our approach creates a link between group fairness and different theories of justice, it does not cover theories of distributive justice that are structurally different from the ones we discussed, e.g., Nozick's entitlement theory \cite{nozick1974anarchy}. It is unclear how such theories could be represented in formalized fairness criteria.

\paragraph{Utility in practice}
While we showcased a simplified approach for specifying utility matrices in Section~\ref{sec:example}, we recognize that defining a utility function is difficult in practice \cite{sen1985standard, elkan2001}.
Moreover, we only presented a utility function that is linear in $Y$ and $D$. Our framework allows for more complex utility functions, but these are even harder to define.
We describe how utility functions can be defined through the lens of a simplified medical example.
However, determining how to quantify the utility of decisions in general (i.e., using a clearly defined guideline that is applicable in any application context, which might require an empirical approach), falls outside the scope of this paper.
Another limitation is that we only proposed simple metrics derived from the utility matrix but no combination of these (e.g., separation as the combination of parity in true positive rates and false positive rates). While we could represent these combined metrics in our framework, it is again not obvious what the best way to do so is. Here, we refer to \cite{fairmlbook} to see how information theory's concept of mutual information can be used to represent separation and sufficiency.

\section{Conclusion}

In this paper, we have proposed a novel generalized definition of group fairness that is based on a comprehensive framework that unifies and extends existing work on what can broadly be described as ``group fairness''.
As part of this, we have also suggested a new definition of group fairness as a category of metrics that are concerned with the just distribution of utility among relevant groups.
Our framework consists of four components: (1) utility of the decision subjects, (2) relevant groups to compare, (3) claim differentiator to derive subgroups to compare that matter, and (4) patterns for a just distribution of utility.
These components form a lens through which we can interpret existing fairness metrics.
The main benefits of our framework are that it allows us to decode the normative choices hidden in fairness criteria and that it yields a structured way of creating unique and context-sensitive fairness criteria.
Using a simple example, we showed that for different versions of prediction-based decision making systems, our approach can determine the fairest solution, according to the chosen normative choices. 
However, the question of how a fair solution can be achieved optimally remains open.
More research is needed to incorporate our novel understanding of group fairness into automated decision making systems, for example, using pre-processing~\cite{Kamiran2009Classifying,kamiran2012dataprocessing}, in-processing~\cite{kamishima2011fairness,zafar2017fairness,menon2018cost}, or post-processing techniques~\cite{hardt2016equality,corbett2017algorithmic,baumann2022sufficiency}.

\begin{acks}
We thank the other members of our project and colleagues (Eleonora Viganò, Ulrich Leicht-Deobald, Serhiy Kandul, Markus Christen, Anikó Hannák, Nicolò Pagan, Stefania Ionescu, Aleksandra Urman, Leonore Röseler, Azza Bouleimen, and Egwuchukwu Ani) for their continuous feedback on the approach presented in this paper.
We also thank participants of our algorithmic fairness workshop at the Applied Machine Learning Days (AMLD) at École polytechnique fédérale de Lausanne (EPFL) in Switzerland and the participants of the course ``Informatics, Ethics and Society'' at the University of Zurich for critical discussions.
This work was supported by the National Research Programme ``Digital Transformation'' (NRP 77) of the Swiss National Science Foundation (SNSF) -- grant number 187473 -- and by Innosuisse -- grant number 44692.1 IP-SBM. Michele Loi was supported by the European Union's Horizon 2020 research and innovation program under the Marie Sklodowska-Curie grant agreement No 898322.
\end{acks}
\bibliographystyle{ACM-Reference-Format}
\bibliography{references}

\appendix

\section{Standard group fairness criteria}
\label{sec:appendix-group-fairness}

Here, we briefly introduce the most discussed group fairness criteria.
Table~\ref{tab:group-fairness-criteria} list the parity requirements associated with these criteria.
\textit{Statistical parity} demands that the share of positive decisions is equal between socio-demographic groups (defined by the sensitive attribute $A=\{0,1\}$)~\cite{dwork2012fairness} -- this is only required for a set of so-called legitimate attributes $l \in L$ for the criterion \textit{conditional statistical parity}~\cite{corbett2017algorithmic}.
\textit{Equality of opportunity}, similarly, demands equal shares of positive decisions between socio-demographic groups, but only for those whose target variable is positive ($Y=1$)~\cite{hardt2016equality} -- thus, it is sometimes also referred to as true positive rate (TPR) parity.
\textit{Equalized odds} -- sometimes also called \textit{separation} -- requires both equality of opportunity and FPR parity (which is similar to equality of opportunity, however, it is limited to individuals of type $Y=0$).
In contrast, \textit{predictive parity} demands equal shares of individuals of type $Y=1$ across socio-demographic groups, but only for those who received a positive decision $D=1$~\cite{baumann2022sufficiency} -- thus, it is sometimes also referred to as positive predictive value (PPV) parity.
\textit{Sufficiency} requires both PPV parity and false omission rate (FOR) parity (which is similar to PPV parity, however, it is limited to individuals who received a negative decision $D=0$).

\begin{table*}[ht!]
\centering
\caption{Standard group fairness criteria}
\label{tab:group-fairness-criteria}
\begin{tabular}{ll}
\toprule
\textbf{Fairness criterion}      & \textbf{Parity requirement}          \\
\midrule
Statistical parity               & $P(D=1|A=0)=P(D=1|A=1)$           \\ \hline
Conditional statistical parity   & $P(D=1|L=l, A=0)=P(D=1|L=l, A=1)$     \\ \hline
Equality of opportunity          & $P(D=1|Y=1, A=0)=P(D=1|Y=1, A=1)$ \\ \hline
False positive rate parity       & $P(D=1|Y=0, A=0)=P(D=1|Y=0, A=1)$ \\ \hline
Equalized odds                   & $P(D=1|Y=y, A=0)=P(D=1|Y=y, A=1)$, for $y\in\{0,1\}$ \\ \hline
Predictive parity                & $P(Y=1|D=1, A=0)=P(Y=1|D=1, A=1)$ \\ \hline
False omission rate parity       & $P(Y=1|D=0, A=0)=P(Y=1|D=0, A=1)$ \\ \hline
Sufficiency                      & $P(Y=1|D=d, A=0)=P(Y=1|D=d, A=1)$, for $d\in\{0,1\}$ \\
\bottomrule
\end{tabular}
\end{table*}

\section{Mapping standard group fairness criteria to our utility-based approach}
\label{sec:appendix}

\renewcommand{\theequation}{B.\arabic{equation}}

\subsection{Omitted proofs}
\label{appendix:omitted_proofs}

\subsubsection{Proof of Proposition~\ref{prop:statistical_parity_as_utility}}
\label{proof:statistical_parity}

Recall that the utility-based fairness following the pattern of egalitarianism requires equal expected utilities between groups:
\begin{equation}
    E(U_{DS}|J=j,A=0)=E(U_{DS}|J=j,A=1)
\end{equation}
Since there is no claim differentiator (i.e., $J=\emptyset$), this can be simplified to:
\begin{equation}
    E(U_{DS}|A=0)=E(U_{DS}|A=1)
    \label{eq:appendix_statistical_parity}
\end{equation}
For $w_{11} = w_{10}$ and $w_{01} = w_{00}$, the decision subject utility (see Equation~\ref{eq:utility-ds-individual}) is:
\begin{equation}
u_{DS,i} = w_{0y} + (w_{1y} - w_{0y}) \cdot d_i,
\end{equation}
where $w_{1y}$ denotes the decision subject utility associated with a positive decision ($D=1$) and $w_{0y}$ denotes the decision subject utility associated with a negative decision ($D=0$).
Thus, the expected utility for individuals of group $a$ can be written as:
\begin{equation}
    E(U_{DS}|A=a) = w_{0y} + (w_{1y} - w_{0y}) \cdot P(D=1|A=a).
\label{eq:appendix_statistical_parity_2}
\end{equation}
If the utility weights of all possible outcomes do not depend on the group membership ($w_{dy} \perp a$), and $w_{1y} \neq w_{0y}$\footnote{If $w_{1y} = w_{0y}$, then the utility-based fairness following the pattern of egalitarianism would always be satisfied and the equivalence to statistical parity would not hold.}, then the utility-based fairness following the pattern of egalitarianism (see Equation~\ref{eq:appendix_statistical_parity}) requires:
\begin{equation}
\begin{split}
    w_{0y} + (w_{1y} - w_{0y}) \cdot P(D=1|A=0) &= w_{0y} + (w_{1y} - w_{0y}) \cdot P(D=1|A=1) \\
    \Leftrightarrow (w_{1y} - w_{0y}) \cdot P(D=1|A=0) &= (w_{1y} - w_{0y}) \cdot P(D=1|A=1)\\
    \Leftrightarrow P(D=1|A=0) &= P(D=1|A=1),
\end{split}
\end{equation}
where the last line is identical to statistical parity.

\subsubsection{Proof of Corollary~\ref{cor:partial_fulfillment_of_statistical_parity}}
\label{proof:partial_fulfillment_of_statistical_parity}

Recall that the degree to which egalitarianism is fulfilled is defined as $F_{\text{egalitarianism}}  = |E(U_{DS}|J=j, A=0) - E(U_{DS}|J=j, A=1)|$ (see Equation~\ref{eq:egalitarianism-measurement}).
If the utility weights of all possible outcomes do not depend on the group membership ($w_{dy} \perp a$), and $w_{11} = w_{10} \neq w_{01} = w_{00}$ (i.e., $w_{1y} \neq w_{0y}$), $J=\emptyset$, this can be written as (see Equations~\ref{eq:appendix_statistical_parity} and~\ref{eq:appendix_statistical_parity_2}):
\begin{equation}
\begin{split}
    F_{\text{egalitarianism}} = & \; | \left( w_{0y} + (w_{1y} - w_{0y}) \cdot P(D=1|A=0) \right) \\
    & \; - \left( w_{0y} + (w_{1y} - w_{0y}) \cdot P(D=1|A=1) \right) | \\
    = & \; | \left( (w_{1y} - w_{0y}) \cdot P(D=1|A=0) \right) - \left( (w_{1y} - w_{0y}) \cdot P(D=1|A=1) \right) | \\
    = & \; | (w_{1y} - w_{0y}) \cdot \left( P(D=1|A=0) - P(D=1|A=1) \right) |
\end{split}
\end{equation}
where the last line corresponds to a multiplication of $|w_{1y} - w_{0y}|$ with the degree to which statistical parity is fulfilled.

\subsubsection{Proof of Proposition~\ref{prop:EOP_as_utility}}
\label{proof:EOP_as_utility}

Recall that the utility-based fairness following the pattern of egalitarianism requires equal expected utilities between groups:
\begin{equation}
    E(U_{DS}|J=j,A=0)=E(U_{DS}|J=j,A=1)
\end{equation}
Since the claim differentiator is the same as the attribute $Y=1$, i.e., $J=Y$ and the only morally relevant value of $Y$ is $1$ (i.e., $j \in \{1\}$), this can be simplified to:
\begin{equation}
    E(U_{DS}|Y=1,A=0)=E(U_{DS}|Y=1,A=1)
    \label{eq:appendix_equality_of_opportunity}
\end{equation}
For $y_i = 1$, the decision subject utility (see Equation~\ref{eq:utility-ds-individual}) is:
\begin{equation}
u_{DS,i} = w_{01} + (w_{11} - w_{01}) \cdot d_i.
\end{equation}
Thus, the expected utility for individuals of type $Y=1$ in group $a$ can be written as:
\begin{equation}
    E(U_{DS}|Y=1,A=a) = w_{01} + (w_{11} - w_{01}) \cdot P(D=1|Y=1,A=a).
    \label{eq:appendix_equality_of_opportunity_2}
\end{equation}
If $w_{11}$ and $w_{01}$ do not depend on the group membership ($w_{d1} \perp a$), and $w_{11} \neq w_{01}$\footnote{If $w_{11} = w_{01}$, then the utility-based fairness following the pattern of egalitarianism would always be satisfied and the equivalence to equality of opportunity would not hold.}, then the utility-based fairness following the pattern of egalitarianism (see Equation~\ref{eq:appendix_equality_of_opportunity}) requires:
\begin{equation}
\begin{split}
    w_{01} + (w_{11} - w_{01}) \cdot P(D=1|Y=1,A=0) &= w_{01} + (w_{11} - w_{01}) \cdot P(D=1|Y=1,A=1) \\
    \Leftrightarrow (w_{11} - w_{01}) \cdot P(D=1|Y=1,A=0) &= (w_{11} - w_{01}) \cdot P(D=1|Y=1,A=1)\\
    \Leftrightarrow P(D=1|Y=1,A=0) &= P(D=1|Y=1,A=1),
\end{split}
\end{equation}
where the last line is identical to equality of opportunity.

\subsubsection{Proof of Proposition~\ref{prop:predictive_parity_as_utility}}
\label{proof:predictive_parity_as_utility}

Recall that the utility-based fairness following the pattern of egalitarianism requires equal expected utilities between groups:
\begin{equation}
    E(U_{DS}|J=j,A=0)=E(U_{DS}|J=j,A=1)
\end{equation}
Since the claim differentiator is the same as the decision $D=1$, i.e., $J=D$ and the only morally relevant value of $D$ is $1$ (i.e., $j \in \{1\}$), this can be simplified to:
\begin{equation}
    E(U_{DS}|D=1,A=0)=E(U_{DS}|D=1,A=1)
    \label{eq:appendix_predictive_parity}
\end{equation}
For $d_i = 1$, the decision subject utility (see Equation~\ref{eq:utility-ds-individual}) is:
\begin{equation}
u_{DS,i} = w_{10} + (w_{11} - w_{10}) \cdot y_i.
\end{equation}
Thus, the expected utility for individuals in group $a$ that are assigned the decision $D=1$ can be written as:
\begin{equation}
    E(U_{DS}|D=1,A=a) = w_{10} + (w_{11} - w_{10}) \cdot P(Y=1|D=1,A=a).
    \label{eq:appendix_predictive_parity_2}
\end{equation}
If $w_{11}$ and $w_{10}$ do not depend on the group membership ($w_{1y} \perp a$), and $w_{11} \neq w_{10}$\footnote{If $w_{11} = w_{10}$, then the utility-based fairness following the pattern of egalitarianism would always be satisfied and the equivalence to predictive parity would not hold.}, then the utility-based fairness following the pattern of egalitarianism (see Equation~\ref{eq:appendix_predictive_parity}) requires:
\begin{equation}
\begin{split}
    w_{10} + (w_{11} - w_{10}) \cdot P(Y=1|D=1,A=0) &= w_{10} + (w_{11} - w_{10}) \cdot P(Y=1|D=1,A=1) \\
    \Leftrightarrow (w_{11} - w_{10}) \cdot P(Y=1|D=1,A=0) &= (w_{11} - w_{10}) \cdot P(Y=1|D=1,A=1)\\
    \Leftrightarrow P(Y=1|D=1,A=0) &= P(Y=1|D=1,A=1),
\end{split}
\end{equation}
where the last line is identical to predictive parity.

\subsection{Additional corollaries}
\label{app:Additional_corollaries}

Let us consider the partial fulfillment of equality of opportunity, following Proposition~\ref{prop:EOP_as_utility}.
As is the case for statistical parity, there are differences when looking at the degree to which the two notions of fairness are fulfilled (equality of opportunity and the utility-based fairness under the conditions specified in Proposition~\ref{prop:EOP_as_utility})
\begin{corollary}[Partial fulfillment of equality of opportunity in terms of utility-based fairness]
    Suppose that the degree to which equality of opportunity is fulfilled is defined as the absolute difference in decision ratios for individuals of type $Y=1$ across groups, i.e., $|P(D=1|Y=1,A=0) - P(D=1|Y=1, A=1)|$.
    If $w_{11}$ and $w_{01}$ do not depend on the group membership ($w_{d1} \perp a$), $w_{11} \neq w_{01}$, $J=Y$, and $j \in \{1\}$, then the degree to which egalitarianism is fulfilled is equivalent to the degree to which equality of opportunity is fulfilled, multiplied by $|(w_{11} - w_{01})|$.
    \label{cor:partial_fulfillment_of_equality_of_opportunity}
\end{corollary}
\begin{proof}
Recall that the degree to which egalitarianism is fulfilled is defined as $F_{\text{egalitarianism}}  = |E(U_{DS}|J=j, A=0) - E(U_{DS}|J=j, A=1)|$ (see Equation~\ref{eq:egalitarianism-measurement}).
If $w_{11}$ and $w_{01}$ do not depend on the group membership ($w_{d1} \perp a$), $w_{11} \neq w_{01}$, $J=Y$, and $j \in \{1\}$, this can be written as (see Equations~\ref{eq:appendix_equality_of_opportunity} and~\ref{eq:appendix_equality_of_opportunity_2}):
\begin{equation}
\begin{split}
    F_{\text{egalitarianism}} = & \; | \left( w_{01} + (w_{11} - w_{01}) \cdot P(D=1|Y=1,A=0) \right) \\
    & \; - \left( w_{01} + (w_{11} - w_{01}) \cdot P(D=1|Y=1,A=1) \right) | \\
    = & \; | \left( (w_{11} - w_{01}) \cdot P(D=1|Y=1,A=0) \right) \\
    & \; - \left( (w_{11} - w_{01}) \cdot P(D=1|Y=1,A=1) \right) | \\
    = & \; | (w_{11} - w_{01}) \cdot \left( P(D=1|Y=1,A=0) - P(D=1|Y=1,A=1) \right) |
\end{split}
\end{equation}
where the last line corresponds to a multiplication of $|w_{11} - w_{01}|$ with the degree to which equality of opportunity is fulfilled.
\end{proof}

As is the case for the other group fairness criteria, there are differences regarding the degree to which the two fairness notions of predictive parity and the utility-based fairness under the conditions specified in Proposition~\ref{prop:predictive_parity_as_utility} are fulfilled:
\begin{corollary}[Partial fulfillment of predictive parity in terms of utility-based fairness]
    Suppose that the degree to which predictive parity is fulfilled is defined as the absolute difference in the ratio of individuals that are of type $Y=1$ among all those that are assigned the decision $D=1$ across groups, i.e., $|P(Y=1|D=1,A=0) - P(Y=1|D=1, A=1)|$.
    If $w_{11}$ and $w_{10}$ do not depend on the group membership ($w_{1y} \perp a$), $w_{11} \neq w_{10}$, $J=D$, and $j \in \{1\}$, then the degree to which egalitarianism is fulfilled is equivalent to the degree to which predictive parity is fulfilled, multiplied by $|w_{11} - w_{10}|$.
    \label{cor:partial_fulfillment_of_predictive_parity}
\end{corollary}
\begin{proof}
Recall that the degree to which egalitarianism is fulfilled is defined as $F_{\text{egalitarianism}}  = |E(U_{DS}|J=j, A=0) - E(U_{DS}|J=j, A=1)|$ (see Equation~\ref{eq:egalitarianism-measurement}).
If $w_{11}$ and $w_{10}$ do not depend on the group membership ($w_{1y} \perp a$), $w_{11} \neq w_{10}$, $J=D$, and $j \in \{1\}$, this can be written as (see Equations~\ref{eq:appendix_predictive_parity} and~\ref{eq:appendix_predictive_parity_2}):
\begin{equation}
\begin{split}
    F_{\text{egalitarianism}} = & \; | \left( w_{10} + (w_{11} - w_{10}) \cdot P(Y=1|D=1,A=0) \right) \\
    & \; - \left( w_{10} + (w_{11} - w_{10}) \cdot P(Y=1|D=1,A=1) \right) | \\
    = & \; | \left( (w_{11} - w_{10}) \cdot P(Y=1|D=1,A=0) \right) \\
    & \; - \left( (w_{11} - w_{10}) \cdot P(Y=1|D=1,A=1) \right) | \\
    = & \; | (w_{11} - w_{10}) \cdot \left( P(Y=1|D=1,A=0) - P(Y=1|D=1,A=1) \right) |
\end{split}
\end{equation}
where the last line corresponds to a multiplication of $|w_{11} - w_{10}|$ with the degree to which predictive parity is fulfilled.
\end{proof}

\subsection{Mapping to other group fairness criteria}
\label{appendix:additional_results}

In Section~\ref{sec:relation-group-fairness}, we mapped our utility-based approach to the three group fairness criteria statistical parity, equality of opportunity, and predictive parity.
Here, we additionally show under which conditions our utility-based approach is equivalent to other group fairness criteria: conditional statistical parity, false positive rate parity, equalized odds, false omission rate parity, and sufficiency.

\subsubsection{Conditional statistical parity}

Conditional statistical parity is defined as $P(D=1|L=l, A=0)=P(D=1|L=l, A=1)$, where $L$ is what~\cite{corbett2017algorithmic} refer to as the \textit{legitimate} attributes.
Thus, conditional statistical parity requires equality of acceptance rates across all subgroups in $A=0$ and $A=1$ who are equal in their value $l$ for $L$, where $L$ can be any (combination of) feature(s) besides $D$ and $A$.
\begin{proposition}[Conditional statistical parity as utility-based fairness]
    If the utility weights of all possible outcomes do not depend on the group membership ($w_{dy} \perp a$), and $w_{11} = w_{10} \neq w_{01} = w_{00}$, then the egalitarian pattern fairness condition with $J=L$ is equivalent to conditional statistical parity.
\label{prop:conditional_statistical_parity_as_utility}
\end{proposition}
The proof of Proposition~\ref{prop:conditional_statistical_parity_as_utility} is similar to the one of Proposition~\ref{prop:statistical_parity_as_utility}.

Under these conditions, the degree to which $F_{\text{egalitarianism}}$ is fulfilled is equivalent to the degree to which conditional statistical parity is fulfilled, multiplied by $|w_{1y} - w_{0y}|$.
This could easily be proved -- similar to the proof of Corollary~\ref{cor:partial_fulfillment_of_statistical_parity} but with the conditions of the utility-based fairness stated in Proposition~\ref{prop:conditional_statistical_parity_as_utility}.

\subsubsection{False positive rate (FPR) parity}

FPR parity (also called predictive equality~\cite{corbett2017algorithmic}) is defined as $P(D=1|Y=0, A=0)=P(D=1|Y=0, A=1)$, i.e., it requires parity of false positive rates (FPR) across groups $a \in A$.
\begin{proposition}[FPR parity as utility-based fairness]
    If $w_{10}$ and $w_{00}$ do not depend on the group membership ($w_{d0} \perp a$), and $w_{10} \neq w_{00}$, then the egalitarian pattern fairness condition with $J=Y$ and $j \in \{0\}$ is equivalent to FPR parity.
\label{prop:FPR_parity_as_utility}
\end{proposition}
For $y_i = 0$, the decision subject utility (see Equation~\ref{eq:utility-ds-individual}) is:
\begin{equation}
u_{DS,i} = w_{00} + (w_{10} - w_{00}) \cdot d_i.
\end{equation}
Thus, the expected utility for individuals of type $Y=0$ in group $a$ can be written as:
\begin{equation}
    E(U_{DS}|Y=0,A=a) = w_{00} + (w_{10} - w_{00}) \cdot P(D=1|Y=0,A=a).
\end{equation}
Hence, we simply require the utility weights $w_{10}$ and $w_{00}$ to be unequal and independent of $a$.
Then, the proof of Proposition~\ref{prop:FPR_parity_as_utility} is similar to the one of Proposition~\ref{prop:EOP_as_utility}.

If $w_{10}$ and $w_{00}$ do not depend on the group membership ($w_{d0} \perp a$), and $w_{10} \neq w_{00}$, then the degree to which $F_{\text{egalitarianism}}$ is fulfilled is equivalent to the degree to which FPR parity is fulfilled, multiplied by $|w_{10} - w_{00}|$.
This could easily be proved -- similar to the proof of Corollary~\ref{cor:partial_fulfillment_of_equality_of_opportunity}.

\subsubsection{Equalized odds}

Equalized odds (sometimes also referred to as separation~\cite{fairmlbook}) is defined as $P(D=1|Y=y, A=0)=P(D=1|Y=y, A=1)$, for $y\in\{0,1\}$.
\begin{proposition}[Equalized odds as utility-based fairness]
    If the utility weights of all possible outcomes do not depend on the group membership ($w_{dy} \perp a$), $w_{11} \neq w_{01}$, and $w_{10} \neq w_{00}$, then the egalitarian pattern fairness condition with $J=Y$ and $j \in \{0,1\}$ is equivalent to equalized odds.
\label{prop:equalized_odds_as_utility}
\end{proposition}
The conditions under which the utility-based fairness criteria is equivalent is shown separately for equality of opportunity (see Proposition~\ref{prop:EOP_as_utility}) and FPR parity (see Proposition~\ref{prop:FPR_parity_as_utility}).
Since equalized odds requires equality of opportunity and FPR parity, the the conditions for both fairness criteria must be met (i.e., $w_{dy} \perp a$), $w_{11} \neq w_{01}$, $w_{10} \neq w_{00}$, $J=Y$, and $j \in \{0,1\}$), so that the utility-based fairness constraint is equivalent to equalized odds.

\subsubsection{False omission rate (FOR) parity}

FOR parity is defined as $P(Y=1|D=0, A=0)=P(Y=1|D=0, A=1)$, i.e., it requires parity of false omission rates (FOR) across groups $a \in A$.
\begin{proposition}[FOR parity as utility-based fairness]
    If $w_{01}$ and $w_{00}$ do not depend on the group membership ($w_{0y} \perp a$), and $w_{01} \neq w_{00}$, then the egalitarian pattern fairness condition with $J=D$, and $j \in \{0\}$ is equivalent to FOR parity.
\label{prop:FOR_parity_as_utility}
\end{proposition}
For $d_i = 0$, the decision subject utility (see Equation~\ref{eq:utility-ds-individual}) is:
\begin{equation}
u_{DS,i} = w_{00} + (w_{01} - w_{00}) \cdot y_i.
\end{equation}
Thus, the expected utility for individuals in group $a$ that are assigned the decision $D=0$ can be written as:
\begin{equation}
    E(U_{DS}|D=0,A=a) = w_{00} + (w_{01} - w_{00}) \cdot P(Y=1|D=0,A=a).
\end{equation}
Hence, we simply require the utility weights $w_{01}$ and $w_{00}$ to be unequal and independent of $a$.
Then, the proof of Proposition~\ref{prop:FOR_parity_as_utility} is similar to the one of Proposition~\ref{prop:predictive_parity_as_utility}.

If $w_{01}$ and $w_{00}$ do not depend on the group membership ($w_{0y} \perp a$), and $w_{01} \neq w_{00}$, then the degree to which $F_{\text{egalitarianism}}$ is fulfilled is equivalent to the degree to which FoR parity is fulfilled, multiplied by $|w_{01} - w_{00}|$.
This could easily be proved -- similar to the proof of Corollary~\ref{cor:partial_fulfillment_of_predictive_parity}.

\subsubsection{Sufficiency}

Sufficiency is defined as $P(Y=1|D=d, A=0)=P(Y=1|D=d, A=1)$, for $d\in\{0,1\}$~\cite{fairmlbook}.
\begin{proposition}[Sufficiency as utility-based fairness]
    If the utility weights of all possible outcomes do not depend on the group membership ($w_{dy} \perp a$), $w_{11} \neq w_{10}$, and $w_{01} \neq w_{00}$, then the egalitarian pattern fairness condition with $J=D$ and $j \in \{0,1\}$ is equivalent to sufficiency.
\label{prop:sufficiency_odds_as_utility}
\end{proposition}
The conditions under which the utility-based fairness criteria is equivalent is shown separately for predictive parity (see Proposition~\ref{prop:predictive_parity_as_utility}) and FOR parity (see Proposition~\ref{prop:FOR_parity_as_utility}).
Since sufficiency requires predictive parity and FOR parity, the the conditions for both fairness criteria must be met (i.e., $w_{dy} \perp a$), $w_{11} \neq w_{10}$, $w_{01} \neq w_{00}$, $J=D$, and $j \in \{0,1\}$), so that the utility-based fairness constraint is equivalent to sufficiency.

\end{document}